\documentstyle[aps,epsfig,prl,twocolumn]{revtex}
\draft
\begin{document}

\twocolumn[\hsize\textwidth\columnwidth\hsize\csname
@twocolumnfalse\endcsname
\title{Collective Singlet Excitations and Evolution of
Raman Spectral Weights in the 2D Spin Dimer Compound
SrCu$_2$(BO$_3$)$_2$}
\author{P. Lemmens, M. Grove, M. Fischer, G. G\"untherodt}
\address{2. Physikalisches Institut, RWTH Aachen, 52056 Aachen, Germany}
\author{Valeri N. Kotov}
\address{Department of Physics, University of  Florida, Gainesville, FL
32611-8440, USA}
\author{H. Kageyama, K. Onizuka, Y. Ueda}
\address{Institute for Solid State Physics,
University of Tokyo, Roppongi 7-22-1, Tokyo 106-8666, Japan}
\maketitle
\begin{abstract}
Raman light scattering of the two-dimensional quantum spin system
SrCu$_2$(BO$_3$)$_2$ shows a rich structure in the magnetic
excitation spectrum, including several well-defined bound state
modes at low temperature, and a scattering continuum and
quasielastic light scattering contributions at high temperature.
The key to the understanding of the unique features of
SrCu$_2$(BO$_3$)$_2$ is the presence of  strong interactions
between  well-localized triplet excitations in the network of
orthogonal spin dimers realized in this compound.
\end{abstract}




]

\narrowtext


Low-dimensional quantum spin systems with a quantum disordered
ground state and a finite spin gap
form a subject which is of both fundamental and applied interest
to the physics community. This is due to the fascinating and
diverse physics of the spin-liquid state, allowing to address
issues related to the nature of quasiparticle excitations and the
role of strong interaction effects. It is known that a gapped
singlet ground state is realized in one dimension (1D) in
dimerized or frustrated spin chains, and in even-leg spin
ladders. Of even greater interest is the existence of a gap in
two-dimensional (2D) spin systems, because of its potential
relevance for the description of high-temperature
superconductivity. Unfortunately, only very few 2D systems, such
as CaV$_4$O$_9$, have been investigated, in which the spin gap is
relatively large and not caused by anisotropies. Recently, the
compound SrCu$_2$(BO$_3$)$_2$ with a layered structure was
identified as a 2D S=1/2 Heisenberg system with a unique exchange
topology \cite{kageyama,smith} leading to an exact dimer ground
state, thus providing the  opportunity to study  the excitation
spectrum of such a model quantum many-body system.

SrCu$_2$(BO$_3$)$_2$ has a tetragonal unit cell (D$_{2d}$) with
Cu$^{2+}$ ions that carry a localized spin S=1/2. The spins form
dimers which consist of neighboring pairs of planar rectangular
CuO$_4$ plaquettes with a Cu-Cu distance of 2.9~{\AA}. The strength of
the antiferromagnetic intradimer exchange coupling is estimated to
be  J$_1$=100~K \cite{miyahara}. The spin dimers are connected
orthogonally by a triangular BO$_3$ unit. The distance between
next-nearest Cu neighbors is 5.1~{\AA}, and the interdimer exchange
J$_2$$\approx$68~K \cite{miyahara}. The orthogonal arrangement of
dimers thus represents a 2D frustrated quantum spin system
\cite{kageyama,miyahara}. A sketch of the Cu-dimers is shown in
the inset of Fig.1. The ground state in this exchange topology
depends critically on the ratio J$_2$/J$_1$
\cite{miyahara,weihong,goetz}. For small  J$_2$ the system
consists of isolated dimers and the  ground state is a product of
singlets, while for  small J$_1$ the model can be mapped on a 2D
square lattice of spins and a N\'eel-ordered state is expected.
It has been shown that the critical ratio of coupling constants
that separates a gapfull and a gapless state is  $J_2/J_1=0.7$
\cite{miyahara,weihong}. The experimental ratio of
$J_2/J_1\approx0.68$ is just below this value, placing this
material in the dimerized phase, close to the  N\'eel boundary.
Thermodynamic measurements support the existence of a dimer ground
state and a spin gap of $\Delta$=34~K
\cite{kageyama,kage-thermodynamic}. The magnetic specific heat
and susceptibility both show maxima (T$_{max}^{\chi}$=15~K,
T$_{max}^{c_p}$=8~K) and a rapid decrease toward lower
temperatures with exponential tails at $T\rightarrow0$.

We have investigated the magnetic excitation spectrum of single
crystals of SrCu$_2$(BO$_3$)$_2$ by Raman scattering experiments.
At low temperatures (T$\ll$$\Delta$) we observe well-defined
low-energy singlet modes which can be interpreted as collective
bound state excitations of two and three elementary triplets. We
demonstrate that the appearance of the collective modes reflects
the strong triplet-triplet interactions present in the system.
The spectral weight evolution for intermediate
(T$\approx$$\Delta$) and high (T$>$$\Delta$) temperatures can be
described by the temperature dependence of the magnetic
susceptibility and the magnetic specific heat.

The samples were investigated in quasi-back\-scat\-tering
geometry with light polarizations in the $ab$-plane of the
freshly cleaved crystal. The ($a'b'$) axes are rotated by
45$^\circ$ within this plane. The experiments used the $\lambda$
= 488-nm excitation line of an Ar$^+$ ion laser and a laser
fluence below 20~W/cm$^2$. The scattered light was analyzed using
a XY-Dilor Raman spectrometer and a back-illuminated CCD
detector. Measurements in a magnetic field were performed in
90$^\circ$ scattering geometry. In the analysis of our results
the  main emphasis has been on the role of interactions and their
influence on the magnetic excitation spectrum. A detailed
analysis of the phonon spectrum will be given elsewhere.

In the low energy region with Raman shifts comparable to the
triplet gap $\Delta$=24cm$^{-1}$, drastic changes and a
large shift of spectral weight occur with decreasing temperature.
This evolution leads finally to the appearance of several new
modes. In Fig.1 Raman spectra of SrCu$_2$(BO$_3$)$_2$ are shown to
illustrate these effects at different temperatures and in two
scattering configurations with light polarizations within the
{\it ab}-plane of the crystal. The spectra in the upper panel (a)
present data in $(ab)$:B$_2$ and the lower panel (b) shows data in
$(a'b')$:B$_1$ scattering configuration. Symmetry components are
denoted with respect to the D$_{2d}$ point group using
$b'$=$a$+$b$.

At low temperatures (T$\ll$$\Delta$) four well-defined modes with
energies $E^{S}$= 30, 46, 56 and 70~cm$^{-1}$ appear. The
dominant intensity of these modes is observed in the $(a'b')$
scattering configuration. They neither split nor shift in an
applied magnetic field up to 6~T as shown in the inset of
Fig.1(b) and therefore are assigned to spin singlets. The only
effect of the magnetic field is observed as a shift of a weak
intensity shoulder at 24~cm$^{-1}$ toward lower frequencies. This
signal corresponds to the elementary spin gap. At higher
temperatures (T$\approx$$\Delta$) all modes are strongly damped.
They are replaced by a continuum of scattering with a center of
gravity near 50~cm$^{-1}$ (Fig.1(a),T=7~K), corresponding roughly
to the energy 2$\Delta$. For even higher temperatures
(T$>$$\Delta$) quasielastic scattering with a Lorentzian spectral
function is detected. The latter two scattering intensities are
observed in the $(ab)$ scattering configuration.

To understand these dramatic changes of the Raman spectra with
temperature we compare them with observations related to the
triplet excitation spectrum and then map the temperature
dependence of the scattering intensity onto the corresponding
thermodynamic data, i.e. the magnetic susceptibility and the
magnetic part of the specific heat. To gain a deeper insight into
the structure of the spectrum we will also identify the
interactions leading to the formation of singlet bound states in
the relevant Heisenberg model at T=0, and present estimates for
their binding energies.

Recent ESR \cite{nojiri} and neutron scattering investigations
\cite{kageyama-neutron} on SrCu$_2$(BO$_3$)$_2$ at low
temperatures observed triplet excitations with a spin gap of
$\Delta$=34~K. This triplet branch has a very small dispersion of
only 2~K pointing to extremely localized excitations
\cite{weihong}. In addition, a second triplet branch
$\Delta'$=55~K with a larger dispersion of 17~K was observed. This
branch can be interpreted as a triplet bound state of two
elementary triplets (see discussion below). Frustration due to
the interdimer coupling J$_2$ can lead to the reduction of the
ratio $\Delta'$/$\Delta$=1.62 below 2 (corresponding to
non-interacting magnons) \cite{kageyama-neutron}.

The four modes that we observe in Raman scattering in
SrCu$_2$(BO$_3$)$_2$ for T$\ll$$\Delta$ are clearly related in
energy and temperature scale to the spin gap of the compound. A
phonon origin of these modes can be ruled out due to their
temperature dependence and the nonexistence of a structural phase
transition below room temperature. On the other hand, in
low-dimensional spin systems with strong triplet-triplet
interactions, well defined modes can appear below the scattering
continuum \cite{uhrig-schulz,bouzerar-PRB,kotov1}. In this case
the light scattering process is better described as spin
conserving ($\Delta$S=0) scattering on singlet bound states. Such
states composed of two elementary triplets have been observed,
e.g., in the low-temperature dimerized phases of $\rm CuGeO_3$ and
$\rm NaV_2O_5$ \cite{cugeo-els,lemmens-navo,lemmens-bound}.

To illustrate the  mechanism of bound state formation at T=0 in
the 2D Heisenberg model $H= \Sigma_{i,j} J_{ij} {\bf S}_{i}\!
\cdot \!{\bf S}_{j}$ with $J_{ij}>0$ defined in the inset of
Fig.1(a) (the Shastry-Sutherland model \cite{shastry}), we have
derived the effective bosonic representation in terms of triplets
(${\bf t}^{\dag}_{i}$), excited above the singlets, formed  on
the stronger ($J_{1}$) bonds:
\begin{eqnarray}
\! \! \! \! H  =  J_{1} \sum_{i} {\bf t}^{\dag}_{i} \cdot {\bf
t}_{i}
+V \sum_{\langle i,j \rangle}( {\bf t}^{\dag}_{i} \times {\bf
t}_{i})\cdot ( {\bf t}^{\dag}_{j} \times {\bf t}_{j})+ \: \: \: \: \nonumber
\\
 \sum_{i \in A}
\left \{ i\Gamma \,
 ({\bf t}^{\dag}_{i} \! \times \! {\bf t}_{i}) \!
\cdot \!  ( {\bf t}^{\dag}_{i+ \hat{x}} \! - \! {\bf
t}^{\dag}_{i-\hat{x}} + \mbox{h.c.}) \right \}
 + \sum_{i \in B} \{\hat{x} \rightarrow \hat{y}\},  \nonumber\\
\end{eqnarray}
where $V=-\Gamma=-J_{2}/2$. The site indices run over the square
lattice formed by the dimers (singlet pairs of spins connected by
$J_{1}$ bonds), and $\langle i,j \rangle$ stands for nearest
neighbors, while sub-lattice A(B) is formed by the vertical
(horizontal) dimers.  An important feature of Eq.(1) is the
absence of quantum fluctuations (i.e. ${\bf t}^{\dag}_{i}\cdot{\bf
t}^{\dag}_{j}$ terms), reflecting the fact that the ground state
is an exact product of singlets.  The triplet excitations
would have been completely localized in the absence of the
$\Gamma$ term in Eq.(1), and hopping appears only in  order
$(\Gamma/J_{1})^{6} \sim (J_{2}/J_{1})^{6}$, leading to a small
bandwidth compared to the gap \cite{miyahara,weihong}.

Next we present  estimates for the energies of the collective,
{\it two-particle excitations}. The zero-momentum two-magnon bound
state with S=0, constructed by exciting two triplets, has the
form: $|\Psi^{S}\rangle = \sum_{i,j}\psi^{S}_{i,j} {\bf
t}^{\dag}_{i} \! \cdot \! {\bf t}^{\dag}_{j} |0\rangle$. Starting
from the limit $J_{2}/J_{1}\ll1$, the  dominant  contribution to
binding comes from the two-particle interaction ($V$ term in
Eq.(1)), with corrections of order $(J_{2}/J_{1})^{2}$ and
higher. This interaction provides an effective attraction between
two triplets. Assuming the triplets are localized the binding
problem can be solved exactly, and the singlet  binding energy
defined as $\epsilon^{S}=2\Delta-E^{S}_{2}$, is
$\epsilon^{S}=Z^{2}J_{2}$. Here $E^{S}_{2}$ is  the energy of the
singlet bound state,  and we have estimated the renormalization
factor $Z$ by evaluating the appropriate lowest order diagrams to
be $Z\approx0.75$ at $J_{2}/J_{1}=0.65$. We have found, by
comparing neutron scattering data  for the one-particle
dispersion from \cite{kageyama-neutron}  with the dispersion
obtained by high-order perturbative expansions \cite{weihong}
that the ratio $J_{2}/J_{1}=0.65$  is consistent with the
experimental results. Indeed, for this ratio the theoretical
bandwidth (BW=difference between the energy at ${\bf k}=(\pi,0)$
and $(0,0)$) and gap are: $BW\approx0.04J_{1},
\Delta\approx0.36J_{1}$ \cite{weihong}, consistent with the
measured values $BW\approx0.3~meV$ and $\Delta\approx3~meV$
\cite{kageyama-neutron}. From the same analysis we estimate
$J_{2}\approx 61K \approx 42~cm^{-1}$. For the higher value of
$J_{2}/J_{1}=0.678$ the bandwidth is approximately twice as large
\cite{weihong} which would make this ratio inconsistent with the
neutron scattering data. Putting everything together we have the
estimate for the singlet binding energy
$\epsilon^{S}=Z^{2}J_{2}\approx24~cm^{-1}$. Our experimental
value for the lowest singlet mode in Fig.1(b) with
$E^{S}_{2}$=30~$cm^{-1}$ is $\epsilon^{S}_{Raman}\approx
18~cm^{-1}$.

The analysis based on the localized picture somewhat
overestimates the binding since corrections of order
$(J_{2}/J_{1})^{2}$ and higher have been neglected. Such
high-order corrections lead to the development of a strong
dispersion as well as  create several two-particle singlet modes
with different energies, classified according to the  group
$D_{2d}$. Detailed calculations have recently been performed
\cite{fukumoto,knetter}, providing quantitatively accurate
results for  the energies in the different symmetry sectors. Only
the modes with $\Gamma_{3}(xy)$ symmetry contribute to the Raman
intensity, and, as  $T\rightarrow0$, their contribution grows for
the ($a'b'$) and vanishes for the ($ab$) geometry \cite{knetter},
consistent with our observations. In addition, the calculated
energies \cite{knetter} are in excellent agreement with the two
lowest ($30~cm^{-1}$ and $46~cm^{-1}$) singlet modes observed in
our experiment, which we therefore interpret as two-particle bound
states.


In SrCu$_2$(BO$_3$)$_2$ the four modes in Fig.1(b) show ratios
$E^{S}$/$\Delta$=1.25 -- 2.9. Two of the well-pronounced peaks
(at $56~cm^{-1}$ and $70~cm^{-1}$) are above the two-particle
threshold ($2\Delta$). The smallness of the  one-particle
bandwidth makes it possible to resolve such higher energy peaks,
since the weight of the scattering background is expected to be
small. The pronounced sharpness of the peaks  indicates that they
can be interpreted as {\it three-particle excitations}, whose
existence is indeed possible due to the localized nature of the
states. A three-particle singlet bound state  can be constructed
as: $|\Phi^{S}\rangle = \sum_{i,j,k}\phi^{S}_{i,j,k} ({\bf
t}^{\dag}_{i} \! \times \! {\bf t}^{\dag}_{j})\cdot {\bf
t}^{\dag}_{k} |0\rangle$, and its energy on the same level of
approximation as discussed above (localized triplets) is:
$E^{S}_{3}=3\Delta - Z^{3}J_{2} \approx 54~cm^{-1}$, in very good
agreement with the energy of the mode at $56~cm^{-1}$. The
appearance of additional levels (as well as finite dispersion) at
higher order is also expected.  The total Raman scattering
intensity, reflecting the presence of two- and three- particle
singlets with spectral weights $I_{2}$ and $I_{3}$, is:
$I_{B}(\omega) = I_{2}~\delta(\omega -
E^{S}_{2})~+~I_{3}~\delta(\omega - E^{S}_{3})$. In order to
estimate the ratio of the two intensities we notice that while
light couples directly to the singlet two-magnon state, the
three-magnon state has to be created via the action of the
$\Gamma$ (magnon number non-conserving) term in Eq.(1) (i.e. this
term provides a vertex correction to the Raman operator $R\sim
{\bf t}^{\dag}_{i}\cdot{\bf t}^{\dag}_{j}$). Consequently, we
estimate  at T=0, $I_{3}/I_{2} \propto (J_{2}/J_{1})^{2} \approx
0.4$, in good agreement with the experimental ratio of 0.32 (at
T=1.5~K). Notice that in the hypothetical case $\Gamma=0$, when
the excitations are completely localized,  the three-particle
states do not contribute to the Raman intensity. Thus the
$\Gamma$ term  plays an important role providing both a finite
bandwidth and coupling to higher bound state modes. Let us  also
mention that three-particle bound states have been predicted
theoretically in quasi-1D systems \cite{kotov1}. Our work presents
the first experimental evidence for their existence in a (2D)
dimerized spin system.

The bound state modes experience strong damping with increasing
temperature (see Fig. 1(b)), due to scattering on thermally
excited triplet states. The decrease of the  scattering intensity
$I_{B}$  with temperature is governed by the density of
excitations:
$I_{B}(T) \propto (1- A e^{\frac{-\Delta}{k_B T}}~)~,$
where $A$ is a constant. In Fig.2(a) the integrated intensity of
the bound states is shown together with a fit based on the
equation above ($A$=215.7) and fixing $\Delta$=34~K from
experimental data. The  macroscopic temperature-induced
population of triplet states has a direct effect on the
excitation spectrum. For intermediate temperatures (5-100~K,
typical spectrum shown at 7~K in Fig.1(a)) a broad continuum of
scattering is observed that is of magnetic origin \cite{muthu}.
However, in comparison with a 2D antiferromagnet with a maximum
of the two-magnon continuum at E$_{max}$=2.8~J \cite{chubukov}
its energy is very small. The reason for the maximum appearing at
only 50~cm$^{-1}\approx$2$\Delta$ is the smallness of the
one-particle bandwidth in SrCu$_2$(BO$_3$)$_2$. The temperature
dependence of the two-magnon scattering intensity shows a
pronounced increase in the temperature range where the intensity
of the bound states is dropping (see the open triangles in
Fig.2(b)). At higher temperatures (not shown here) it decreases
again and is then superimposed on the intensity of quasielastic
scattering. In Fig.2(b) we compare the integrated scattering
intensity with the behavior of the magnetic susceptibility
$\chi$(T). Calculations of the latter quantity exist, but their
agreement with experiment is not fully satisfactory
\cite{kageyama,weihong} and consequently we take $\chi$(T)
directly from experiment \cite{kageyama}. Notice that both the
two-magnon continuum intensity and $\chi$(T) exhibit strong
variations on the same temperature scale, even though we are not
aware of a rigorous sum rule relating the two quantities.

Quasielastic scattering connected with fluctuations of the
magnetic energy density is found at high temperatures
(T$>$$\Delta$) in our Raman scattering experiments. It has the
expected Lorentzian spectral function \cite{yamada}. To determine
the evolution as a function of temperature we use the
hydrodynamic form of the respective correlation function
\cite{Hohenberg}, which includes the magnetic specific heat
$c_m(T)$ and the thermal diffusion constant $D_T$. In the high
temperature approximation ($\hbar\omega/k_BT\ll 1$) the result is
\cite{yamada}:
$I(\omega)\propto\frac{k_B}{\hbar}\frac{c_mT^2D_Tk^2}{\omega^2+(D_Tk^2)^2}~$
, where $k$ is the scattering wave-vector. A fit to this equation
can then be used to estimate $c_m(T)$ from the integrated
intensity, scaled by T$^2$. In Fig.2(c) the result of this
procedure is found in good agreement with the measured specific
heat.

In conclusion, we have shown that the low-energy excitation
spectrum of the 2D compound SrCu$_2$(BO$_3$)$_2$ has a rich and
complex structure. Our main result is that at low temperature
(T$\ll$$\Delta$) the spectrum contains several well pronounced
singlet modes which are interpreted as collective two-particle
and novel three-particle sing\-let bound states of strongly
localized triplets. We have demonstrated that since quantum
fluctuations are absent in the ground state (due to the unique
dimer arrangement), the triplet-triplet interactions can lead to
large binding energies and scattering intensities, and
consequently make the collective modes observable. At intermediate
(T$\sim$$\Delta$) and high (T$\gg$$\Delta$) temperatures, where
the quantum (interaction) effects are  not important, the light
scattering is dominated by a two-magnon continuum and
quasielastic contributions. We relate the spectral weight
evolution in these regimes to thermodynamic quantities.

We acknowledge fruitful discussions with C. Pinettes, P.H.M. van
Loosdrecht, G.S. Uhrig, C. Gros, R. Valent\'\i, W. Brenig,  W.
Atkinson, P. Hirschfeld, and D. Tanner. Financial support by
DFG/SFB 341  and NSF Grant DMR-9357474 (V.N.K.) is gratefully
acknowledged.

\vspace{2cm}
\begin{figure}

\centerline{\epsfig{figure=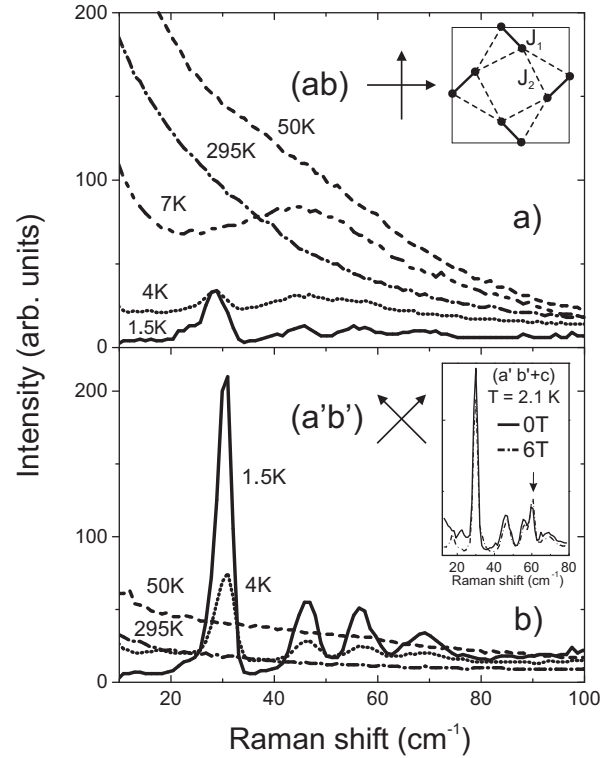,width=8cm}} \vspace{1cm}
\caption{Raman spectra of SrCu$_2$(BO$_3$)$_2$ in two scattering
geometries. The upper inset shows Cu-dimers of 1/4 of the unit cell formed
by J$_1$ (full line) and  J$_2$ (dashed line). The respective polarization
(arrows) of incident and scattered light is given with respect to the
crystallographic axes. The lower inset shows spectra for B=0 (full line)
and 6 Tesla (dotted line) in ($a'~b'+c$) polarization. An additional
A$_g$-phonon is marked by an arrow. }
\end{figure}

\begin{figure}
\centerline{\epsfig{figure=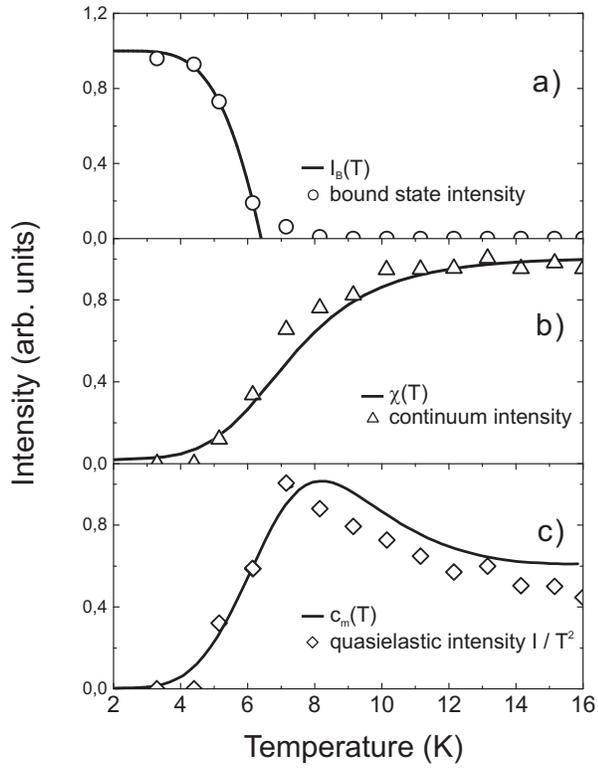,width=8cm}} \vspace{1cm}
\caption{Mapping of Raman scattering intensities (open symbols) on
thermodynamic quantities (lines) as a function of temperature. The
scattering intensities and the thermodynamic data from
Ref.\protect\onlinecite{kageyama,kage-thermodynamic} are normalized to
their maxima.}

\end{figure}

\end{document}